\documentclass[12pt]{article}
\usepackage{amssymb,amsmath,epsfig}

\begin{document}

\title{\bf Anisotropic Universe Models in Brans-Dicke Theory}

\author{M. Sharif \thanks{msharif.math@pu.edu.pk} and Saira Waheed
\thanks{smathematics@hotmail.com}\\
Department of Mathematics, University of the Punjab,\\
Quaid-e-Azam Campus, Lahore-54590, Pakistan.}

\date{}

\maketitle
\begin{abstract}
This paper is devoted to study Bianchi type I cosmological model in
Brans-Dicke theory with self-interacting potential by using perfect,
anisotropic and magnetized anisotropic fluids. We assume that the
expansion scalar is proportional to the shear scalar and also take
power law ansatz for scalar field. The physical behavior of the
resulting models are discussed through different parameters. We
conclude that in contrary to the universe model, the anisotropic
fluid approaches to isotropy at later times in all cases which is
consistent with observational data.
\end{abstract}

\section{Introduction}

Many astronomical experiments and recent cosmological observations
\cite{1} indicate accelerated expansion of our universe. This
expansion is believed by dark energy (DE), a cryptic exotic matter
having large negative pressure that violates the strong energy
condition. In radiation dominated era, the nucleosynthesis
scenario elicits the decelerated expansion of the universe in its
early phase. To understand the nature of DE, many cosmological
models like Chaplygin gas, phantom, quintessence and cosmological
constant etc. have been proposed \cite{2}. The modified theories
of gravity like $f(R)$ gravity, Gauss-Bonnet theory, higher
dimensional theories of gravity, scalar tensor theories etc. have
also been suggested \cite{3}. Brans-Dicke (BD) theory of gravity
is one of the most attractive scalar tensor theories due to its
vast cosmological implications \cite{4}. The varying gravitational
constant ($\frac{1}{\phi}$ acts as gravitational constant), the
non-minimal coupling between the scalar field and geometry,
compatibility with weak equivalence principle, Mach's principle
and Dirac's large number hypothesis are some dominant features of
this theory \cite{5, 6}. The BD parameter should be constrained
$\omega\geq40,000$ for its consistency with the solar system
bounds \cite{7}.

Spatially homogeneous and anisotropic Bianchi type I (BI) model is
used to study the possible effects of anisotropy in the early
universe \cite{8}. Some people \cite{9} have constructed
cosmological models by using anisotropic fluid and BI universe.
Recently, this model has been studied in the presence of binary
mixture of the perfect fluid and the DE \cite{10}. Sharif and
Kausar \cite{11} have discussed dynamics of the universe with
anisotropic fluid and Bianchi models in $f(R)$ gravity. Some exact
BI solutions have also been investigated in this modified theory
\cite{12}.

In this paper, we construct solutions of the field equations for
BI universe model in the presence of different fluids. The paper
is organized as follows. In the next section, we formulate the
field equations of BD theory for BI universe and some general
parameters. Section \textbf{3} provides solution to the field
equations in the presence of perfect fluid and then anisotropic
fluid. The BI cosmological model with magnetized anisotropic fluid
is investigated in section \textbf{4}. A special case, $m=1$, of
the magnetized anisotropic fluid is also discussed. Finally, we
summarize the results in the last section.

\section{Bianchi Type I Field Equations and Some General Parameters}

The BD theory with self-interacting potential is described by the
action \cite{13}
\begin{equation}\label{1}
S=\int d^{4}x\sqrt{-g}[\phi
R-\frac{\omega_{0}}{\phi}\phi^{,\alpha}\phi_{,\alpha}-U(\phi)+L_{m}],\quad\alpha=0,1,2,3,
\end{equation}
where $\omega_{0}$ and $L_{m}$ represent the constant BD parameter
and the matter part of the Lagrangian respectively. Here we have
taken $8\pi G_{0}=c=1$. Using the principle of least action, we
obtain the field equations
\begin{eqnarray}\label{2}
G_{\mu\nu}&=&\frac{\omega_{0}}{\phi^{2}}[\phi_{,\mu}\phi_{,\nu}
-\frac{1}{2}g_{\mu\nu}\phi_{,\alpha}\phi^{,\alpha}]+\frac{1}{\phi}[\phi_{,\mu;\nu}
-g_{\mu\nu}\Box\phi]+\frac{T_{\mu\nu}}{\phi}-g_{\mu\nu}\frac{U(\phi)}{2\phi},\\\label{3}
\Box\phi&=&\frac{T}{3+2\omega_{0}}-\frac{2U(\phi)-\phi\frac{dU(\phi)}{d\phi}}{3+2\omega_{0}}.
\end{eqnarray}
Here $T_{\mu\nu},~T, ~\Box, ~\Delta^{\mu},~U(\phi)$ represent
energy-momentum tensor, its trace, box or d'Alembertian operator
$(\Box=\Delta^{\mu}\Delta_{\mu})$, covariant derivative and the
self-interacting potential respectively. Equation (\ref{3})
represents the Klein Gordon equation or the wave equation for the
scalar field. This theory reduces to general relativity (GR) when
the scalar field is constant and the BD parameter is very large,
i.e., $\omega \rightarrow \infty$ \cite{14}. However this is not
true in general, e.g, the case of exact solutions. It is argued
that this theory goes over to GR only for the non-vanishing trace
of the energy-momentum tensor \cite{15}. For different values of
$\omega$, this theory corresponds to other alternative theories of
gravity. For example, it corresponds to Palatini metric $f(R)$
gravity, the metric $f(R)$ gravity and low energy string theory
action for $\omega=-3/2,~\omega=0$ \cite{16} and $\omega=-1$
\cite{17} respectively.

The BI universe model is given by \cite{18}
\begin{equation}\label{4}
ds^{2}=dt^{2}-A^{2}(t)dx^{2}-B^{2}(t)(dy^{2}+dz^{2}),
\end{equation}
where $A$ and $B$ are the scale factors. This model has one
transverse direction $x$ and two equivalent longitudinal
directions $y$ and $z$. The field equations (\ref{2}) and
(\ref{3}) for the model (\ref{4}) can be written as
\begin{eqnarray}\label{5}
\frac{2\dot{A}\dot{B}}{AB}+\frac{\dot{B}^{2}}{B^{2}}
&=&\frac{T_{00}}{\phi}+\frac{\omega_{0}}{2}
\frac{\dot{\phi}^{2}}{\phi^{2}}-(\frac{\dot{A}}{A}
+2\frac{\dot{B}}{B})\frac{\dot{\phi}}{\phi}+\frac{U(\phi)}{2\phi},\\\label{6}
2\frac{\ddot{B}}{B}+\frac{\dot{B}^{2}}{B^{2}}&=&-\frac{T_{11}}{\phi}
-\frac{\omega_{0}}{2}\frac{\dot{\phi}^{2}}{\phi^{2}}-2\frac{\dot{B}}{B}\frac{\dot{\phi}}{\phi}
-\frac{\ddot{\phi}}{\phi}+\frac{U(\phi)}{2\phi},
\\\label{7}\frac{\ddot{B}}{B}+\frac{\ddot{A}}{A}+\frac{\dot{A}\dot{B}}{AB}
&=&-\frac{T_{22}}{\phi}-\frac{\omega_{0}}{2}
\frac{\dot{\phi}^{2}}{\phi^{2}}-\frac{\ddot{\phi}}{\phi}-(\frac{\dot{A}}{A}
+\frac{\dot{B}}{B})\frac{\dot{\phi}}{\phi}+\frac{U(\phi)}{2\phi}
\end{eqnarray}
and the wave equation is
\begin{equation}\label{8}
\ddot{\phi}+(\frac{\dot{A}}{A}+2\frac{\dot{B}}{B})\dot{\phi}
=\frac{T}{(2\omega_{0}+3)}
-\frac{2U(\phi)-\phi\frac{dU}{d\phi}}{(2\omega_{0}+3)}.
\end{equation}
The corresponding average scale factor $a(t)$, volume $V$ and the
mean Hubble parameter $H$ are
\begin{equation*}
a(t)=(AB^{2})^{1/3},\quad V=a^3(t)=AB^{2},\quad
H(t)=\frac{1}{3}(\frac{\dot{A}}{A}+2\frac{\dot{B}}{B}).
\end{equation*}
The directional Hubble parameters in $x,~y$ and $z$ directions are
given by
\begin{equation}\label{9}
H_{x}=\frac{\dot{A}}{A},\quad H_{y}=H_{z}=\frac{\dot{B}}{B}.
\end{equation}
The anisotropy parameter of expansion $\Delta$ and the deceleration
parameter $q$ are
\begin{equation}\label{10}
\Delta=\frac{1}{3}\sum^{3}_{i=1}(\frac{H_{i}-H}{H})^{2},\quad
q=\frac{d}{dt}(\frac{1}{H})-1.
\end{equation}
The isotropic expansion of the universe can be obtained for
$\Delta=0$. The expansion and shear scalar turn out to be
\begin{eqnarray}\label{11}
\Theta=u^{a}_{;a}=\frac{\dot{A}}{A}+2\frac{\dot{B}}{B},\quad
\sigma=\frac{1}{\sqrt{3}}(\frac{\dot{A}}{A}-\frac{\dot{B}}{B}).
\end{eqnarray}

Since the field equations are highly non-linear, we assume power
law for the scalar field $\phi(t)=\phi_{0}B^{\alpha},~\alpha>0$
for the expanding universe. For a spatially homogeneous metric,
the normal congruence to homogeneous expansion implies that
$\frac{\sigma}{\Theta}$ is constant, i.e., "the expansion scalar
$\Theta$ is proportional to shear scalar $\sigma$" \cite{19}. This
leads to $A=B^{m},~m\neq 1$ for BI model \cite{18}, \cite{20}. It
is worthwhile to mention here that any universe model becomes
isotropic when $t\rightarrow +\infty, ~\Delta\rightarrow 0,~
V\rightarrow +\infty,~\rho>0$ for the diagonal energy-momentum
tensor.

\section{Anisotropic Fluid Model}

In this section, we first explore the BI model with the
energy-momentum tensor of perfect fluid given by
\begin{equation}\label{12}
T^{\mu}_{\nu}=diag[\rho, -\omega\rho, -\omega\rho,-\omega\rho],
\end{equation}
where $\rho$ and  $\omega$ represent the energy density and equation
of state (EoS) parameter respectively. Using this energy-momentum
tensor in the field equations (\ref{2}) and (\ref{3}), it follows
that
\begin{eqnarray}\label{13}
(2m+1)\frac{\dot{B}^{2}}{B^{2}}&=&\frac{\rho}{\phi}+\frac{\omega_{0}}{2}
\frac{\dot{\phi}^{2}}{\phi^{2}}-(m+2)\frac{\dot{B}}{B}\frac{\dot{\phi}}{\phi}
+\frac{U(\phi)}{\phi},\\\label{14}
2\frac{\ddot{B}}{B}+\frac{\dot{B}^{2}}{B^{2}}
&=&-\frac{\omega\rho}{\phi} -\frac{\omega_{0}}{2}
\frac{\dot{\phi}^{2}}{\phi^{2}}-2\frac{\dot{B}}{B}\frac{\dot{\phi}}{\phi}
-\frac{\ddot{\phi}}{\phi}+\frac{U(\phi)}{\phi},\\\nonumber
(m+1)\frac{\ddot{B}}{B}+m^{2}\frac{\dot{B}^{2}}{B^{2}}
&=&-\frac{\omega\rho}{\phi}-\frac{\omega_{0}}{2}
\frac{\dot{\phi}^{2}}{\phi^{2}}-\frac{\ddot{\phi}}{\phi}
-(m+1)\frac{\dot{B}}{B}\frac{\dot{\phi}}{\phi}\\\label{15}
&+&\frac{U(\phi)}{2\phi},\\\label{16}
\ddot{\phi}+(m+2)\frac{\dot{B}}{B}\dot{\phi}
&=&\frac{\rho(1-3\omega)}{(2\omega_{0}+3)}
-\frac{2U(\phi)-\frac{dU}{d\phi}}{(2\omega_{0}+3)},
\end{eqnarray}
where we have used $A=B^{m}$.

The energy conservation equation for such a fluid is
\begin{equation}\label{17}
\dot{\rho}+3H(1+\omega)\rho=0.
\end{equation}
For $0\leq\omega\leq1$, this equation yields
\begin{equation}\label{18}
\rho=\rho_{0}B^{-(m+2)(1+\omega)}.
\end{equation}
Subtracting Eq.(\ref{14}) from (\ref{15}) and using
$\phi=\phi_{0}B^{\alpha}~(\alpha>0)$, we obtain
\begin{equation}\label{19}
B(t)=[(\alpha+m+2)(k_{1}t+k_{2})]^{1/(\alpha+m+2)};\quad m\neq1
\end{equation}
where $k_{1}$ and $k_{2}$ are constants of integration.
Consequently, we have
\begin{equation}\label{20}
A(t)=[(\alpha+m+2)(k_{1}t+k_{2})]^{m/(\alpha+m+2)}.
\end{equation}
Thus the model turns out to be
\begin{eqnarray}\nonumber
ds^{2}&=&dt^{2}-[(\alpha+m+2)(k_{1}t+k_{2})]^{2m/(\alpha+m+2)}dx^{2}\\\label{21}
&-&[(\alpha+m+2)(k_{1}t+k_{2})]^{2/(\alpha+m+2)}(dy^{2}+dz^{2}).
\end{eqnarray}
The corresponding parameters become
\begin{eqnarray}\nonumber
H_{x}&=&mH_{y}=mH_{z}=\frac{\dot{B}}{B}=\frac{mk_{1}}{(\alpha+m+2)(k_{1}t+k_{2})},\\\nonumber
H&=&(\frac{m+2}{3})[\frac{k_{1}}{(\alpha+m+2)(k_{1}t+k_{2})}],\\\nonumber
\Theta&=&\frac{(m+2)k_{1}}{(\alpha+m+2)(k_{1}t+k_{2})},\\\nonumber
\sigma^{2}&=&\frac{(m-1)^{2}}{3}[\frac{k_{1}^{2}}{(\alpha+m+2)^{2}(k_{1}t+k_{2})^{2}}],\\\nonumber
V&=&B^{(m+2)}=[(\alpha+m+2)(k_{1}t+k_{2})]^{\frac{(m+2)}{(\alpha+m+2)}},\\\nonumber
q&=&\frac{d}{dt}(\frac{1}{H})-1=\frac{3\alpha+2m+4}{(m+2)}.
\end{eqnarray}
Since $\alpha,~m>0~(m\neq1)$, we have $q>0$ which yields the
decelerated expansion of the universe. The mean anisotropic
parameter of expansion $(\Delta=\frac{2(m-1)^{2}}{(m+2)^2})$ is
constant.

In order to investigate the accelerated expansion model of the
universe, we take the generalization of the perfect fluid, i.e.,
anisotropic fluid given by
\begin{equation}\label{22}
T^{\nu}_{\mu}=diag[\rho, -p_{x}, -p_{y}, -p_{z}],
\end{equation}
where $\rho$ represents the energy density of the fluid while
$p_{x},~p_{y}$ and $p_{z}$ denote pressures in $x,~y$ and $z$
directions respectively. Equation of state for this fluid is taken
as $p=\omega\rho$, where EoS parameter $\omega$ may not be constant.
By taking the directional EoS parameters
$\omega_{x}=\omega+\delta,~\omega_{y}=\omega+\gamma$ and
$\omega_{z}=\omega+\gamma$ on $x,~y$ and $z$ axes respectively,
Eq.(\ref{22}) can be written as
\begin{equation}\label{23}
T^{\nu}_{\mu}=diag[1, -(\omega+\delta), -(\omega+\gamma),
-(\omega+\gamma)]\rho,
\end{equation}
where $\delta$ denotes deviation from $\omega$ on $x$ axis while
$\gamma$ denotes deviations on $y$ and $z$ axis. Equation
(\ref{23}) with $\delta=0=\gamma$ corresponds to the
energy-momentum tensor for isotropic fluid. The energy
conservation equation for the anisotropic fluid yields
\begin{eqnarray}\label{24}
\dot{\rho}+(1+\omega)(\frac{\dot{A}}{A}+2\frac{\dot{B}}{B})\rho(t)
+(\delta\frac{\dot{A}}{A}+2\gamma\frac{\dot{B}}{B})\rho(t)=0.
\end{eqnarray}
By decomposing the anisotropic fluid into deviation free and
anisotropy parts, we take anisotropy part equal to zero \cite{18,
21}
\begin{equation}\label{25}
(\delta\frac{\dot{A}}{A}+2\gamma\frac{\dot{B}}{B})\rho(t)=0.
\end{equation}
Since $\rho\neq 0$, this implies that either both the deviation
parameters $\delta(t)$ and $\gamma(t)$ vanish or
$\frac{H_{x}}{H_{y}}=-\frac{2\gamma}{\delta}$. For a more general
solution, we take dimensionless deviation parameters as follows
\cite{21}
\begin{eqnarray}\label{26}
\delta(t)=\frac{2n}{3}\frac{\dot{B}}{B}(\frac{\dot{A}}{A}
+2\frac{\dot{B}}{B})\frac{1}{\rho},\quad
\gamma(t)=-\frac{n}{3}\frac{\dot{A}}{A}(\frac{\dot{A}}{A}
+2\frac{\dot{B}}{B})\frac{1}{\rho},
\end{eqnarray}
where $n$ is a real dimensionless constant which describes the
deviation from EoS parameter.

The field equations for such fluid will be
\begin{eqnarray}\label{27}
(2m+1)\frac{\dot{B}^{2}}{B^{2}}&=&\frac{\rho}{\phi}+\frac{\omega_{0}}{2}
\frac{\dot{\phi}^{2}}{\phi^{2}}-(m+2)\frac{\dot{B}}{B}\frac{\dot{\phi}}{\phi}
+\frac{U(\phi)}{\phi},\\\nonumber
2\frac{\ddot{B}}{B}+\frac{\dot{B}^{2}}{B^{2}}&=&-\frac{(\omega+\delta)\rho}{\phi}
-\frac{\omega_{0}}{2}\frac{\dot{\phi}^{2}}{\phi^{2}}-2\frac{\dot{B}}{B}\frac{\dot{\phi}}{\phi}
-\frac{\ddot{\phi}}{\phi}+\frac{U(\phi)}{\phi},\\\label{28}\\\nonumber
(m+1)\frac{\ddot{B}}{B}+m^{2}\frac{\dot{B}^{2}}{B^{2}}
&=&-\frac{(\omega+\gamma)\rho}{\phi}-\frac{\omega_{0}}{2}
\frac{\dot{\phi}^{2}}{\phi^{2}}-\frac{\ddot{\phi}}{\phi}
-(m+1)\frac{\dot{B}}{B}\frac{\dot{\phi}}{\phi}\\\label{29}&+&\frac{U(\phi)}{2\phi},\\\label{30}
\ddot{\phi}+(m+2)\frac{\dot{B}}{B}\dot{\phi}&=&\frac{\rho(1-3\omega)
-\rho(2\gamma+\delta)}{(2\omega_{0}+3)}-\frac{(2U(\phi)-\frac{dU}{d\phi})}{(2\omega_{0}+3)}.
\end{eqnarray}
Using Eqs.(\ref{26}), (\ref{28}) and (\ref{29}) along with
$\phi=\phi_{0}B^{\alpha}$, it follows that
\begin{equation*}
\frac{\ddot{B}}{B}+(m+1+\alpha)\frac{\dot{B}^{2}}{B^{2}}
-\frac{n(m+2)^{2}\dot{B}^{2}}{3B^{(\alpha+2)}(m-1)\phi_{0}}=0.
\end{equation*}
Integrating twice, we obtain
\begin{equation*}
t+k_{4}=\int
B^{(m+1+\alpha)}e^{-(k_{3}-\frac{n(m+2)^{2}B^{-\alpha}}{3\phi_{0}\alpha(m-1)})}dB,
\end{equation*}
where $k_{3}$ and $k_4$ are integration constants. For $B=T,~
x=X,~y=Y$ and $z=Z$, BI model turns out to be
\begin{equation}\label{31}
ds^{2}=T^{-(m+1+\alpha)}e^{(k_{3}-\frac{n(m+2)^{2}T^{-\alpha}}{3\phi_{0}\alpha(m-1)})}dT^{2}
-T^{2m}dX^{2}-T^{2}(dY^{2}+dZ^{2}).
\end{equation}
Some physical parameters are
\begin{eqnarray}\nonumber
V&=&T^{m+2},\quad \Delta=\frac{2(m-1)^{2}}{(m+2)^2},\\\nonumber
H_{x}&=&mH_{y}=m[2-\frac{nlT^{-\alpha}}{3\alpha\phi_{0}(m-1)}]T^{-(m+1+\alpha)},\\\nonumber
H&=&\frac{(m+2)}{3}[2-\frac{nlT^{-\alpha}}{3\alpha\phi_{0}(m-1)}]T^{-(m+1+\alpha)}\\\nonumber
\Theta&=&3H=(m+2)[2-\frac{nlT^{-\alpha}}{3\alpha\phi_{0}(m-1)}]T^{-(m+1+\alpha)},\\\nonumber
\sigma^2&=&\frac{(m-1)^2}{3}[2-\frac{2nlT^{-\alpha}}{3\alpha\phi_{0}(m-1)}]T^{-2(m+1+\alpha)},\\\nonumber
q&=&-(1-\frac{3}{(m+2)})-\frac{3}{(m+2)}(\frac{nl(m+1+2\alpha)T^{-(m+2+2\alpha)}}{3\alpha\phi_{0}(m-1)}\\\nonumber
&-&2(m+1+\alpha)T^{-(m+2+\alpha)})(2-\frac{2nlT^{-\alpha}}{3\alpha\phi_{0}(m-1)})^{-1}T^{(m+2+\alpha)}.
\end{eqnarray}

Since $\alpha,~m>0~(m\neq1)$, these parameters except the
deceleration parameter, increase with the decrease in $T$ and
approach to zero as $T\rightarrow\infty$. Also, for earlier times,
the volume of the universe is zero while the expansion and shear
scalar turn out to be infinite. For later times, the volume goes
to infinite value while the expansion and shear scalar decrease to
zero. This indicates that the universe expands from zero volume at
infinite rate of expansion. Since the anisotropy parameter of
expansion is constant (it vanishes for $m=1$), therefore the model
does not isotropize for later times. In this case, the
deceleration parameter $q$ is found to be a dynamical quantity and
can be negative for the appropriate values of the constant
parameters. For later times and $m>1$, the deceleration parameter
turns out to be negative.

The self-interacting potential $U$ can be written from Eq.(\ref{27})
as follows
\begin{eqnarray}\nonumber
U(\phi)\approx
U(T)&=&2\phi_{0}((\alpha+2)m-\frac{\omega_{0}\alpha^{2}}{2}+1+2m)e^{(2k_{3}
-\frac{2n(m+2)^{2}}{3\phi_{0}(m-1)\alpha T^{\alpha}})}\\\label{32}
&\times&T^{-(2m+4+\alpha)}-2\rho.
\end{eqnarray}
Equations (\ref{24}) and (\ref{25}) lead to
\begin{equation}\label{33}
\omega=-1-\frac{\frac{d\rho}{dt}}{(m+2)\rho\frac{\dot{B}}{B}}.
\end{equation}
Substituting Eqs.(\ref{32}) and (\ref{33}) in (\ref{30}), we
obtain
\begin{eqnarray}\nonumber
\rho(T)&=&[\phi_{0}\alpha(m+2)((3+2\omega_{0})\alpha(\alpha+m+1)+4(1+2m)
+4\alpha(m+2)\\\nonumber
&-&2\omega_{0}\alpha^{2})-\frac{\phi_{0}\alpha^{2}(m+2)(3+2\omega_{0})}{2}
(\frac{8\alpha(m+2)}{(3\alpha-2(m+2))}+\alpha-2)]\\\nonumber
&\times&[2T^{-(\alpha+2m+4)}(8\alpha(m+2)-(\alpha+2m
+4)(3\alpha-2(m+2)))^{-1}\\\nonumber
&-&2nlT^{-(2\alpha+2m+4)}(3\alpha\phi_{0}(m-1)(8\alpha(m+2)-(\alpha+2m
+4)(3\alpha\\\nonumber
&-&2(m+2))))^{-1}]+\frac{4\alpha(m+2)^{2}n(m-1)}{3}[T^{-(4+2\alpha+2m)}(8\alpha(m+2)\\\nonumber
&-&(2\alpha+2m+4)(3\alpha-2(m+2)))^{-1}-nlT^{-(4+3\alpha+2m)}(3\alpha\phi_{0}(m-1)\\\nonumber
&\times&(8\alpha(m+2)-(3\alpha+2m
+4)(3\alpha-2(m+2))))^{-1}]+\alpha^{2}\phi_{0}\\\nonumber
&\times&\frac{(3+2\omega_{0})(m+2)}{(3\alpha-2(m+2))}
[T^{-(\alpha+2m+4)}-\frac{nlT^{-(2\alpha+2m+4)}}{3\alpha\phi_{0}(m-1)}]\\\label{34}
&+&c_{1}T^{\frac{-8\alpha}{(3\alpha-2(m+2))}},
\end{eqnarray}
where $c_{1}$ is an integration constant. Inserting this value in
Eq.(\ref{32}), one can obtain the corresponding self-interacting
potential. The skewness parameters are given by
\begin{eqnarray}\label{35}
\delta(T)&=&\frac{2n(m+2)(2-\frac{2n(m+2)^{2}T^{-\alpha}}
{3\phi_{0}\alpha(m-1)})T^{-2(m+2+\alpha)}}{3\rho},\\\label{36}
\gamma(T)&=&\frac{-nm(m+2)(2-\frac{2n(m+2)^{2}T^{-\alpha}}{
3\phi_{0}\alpha(m-1)})T^{-2(m+2+\alpha)}}{3\rho}.
\end{eqnarray}

The deviation free EoS parameter (\ref{33}) can be written as
\begin{eqnarray}\nonumber
\omega(T)&=&-1-\frac{1}{\rho(m+2)}[[\phi_{0}\alpha(m+2)((3+
2\omega_{0})\alpha(\alpha+m+1)\\\nonumber
&+&4(1+2m)+4\alpha(m+2)-2\omega_{0}\alpha^{2})-(\alpha-2
+\frac{8\alpha(m+2)}{(3\alpha-2(m+2))}) \\\nonumber
&\times&\frac{\phi_{0}\alpha^{2}(m+2)(3+2\omega_{0})}{2}]
[-2(\alpha+2m+4)T^{-(\alpha+2m+5)}(8\alpha(m+2)\\\nonumber
&-&(\alpha+2m+4)(3\alpha-2(m+2)))^{-1}+2nl(2\alpha+2m+4)
\end{eqnarray}
\begin{eqnarray}\nonumber
&\times&T^{-(2\alpha+2m+5)}(3\alpha\phi_{0}(m-1)(8\alpha(m+2)-(\alpha+2m+4)\\\nonumber
&\times&(3\alpha-2(m+2))))^{-1}]+\frac{4\alpha(m+2)^{2}n(m-1)}{3}[-(4+2\alpha+2m)\\\nonumber
&\times&T^{-(5+2\alpha+2m)}(8\alpha(m+2)-(2\alpha+2m
+4)(3\alpha-2(m+2)))^{-1}+nl\\\nonumber
&\times&(4+3\alpha+2m)T^{-(5+3\alpha+2m)}
(3\alpha\phi_{0}(m-1)(8\alpha(m+2)-(3\alpha+2m\\\nonumber
&+&4)(3\alpha-2(m+2))))^{-1}]+\alpha^{2}\phi_{0}\frac{(3+2\omega_{0})(m+2)}{(3\alpha-2(m+2))}
[-(\alpha+2m+4)\\\nonumber
&\times&T^{-(\alpha+2m+5)}+\frac{nl(2\alpha+2m+4)T^{-(2\alpha+2m+5)}}
{3\alpha\phi_{0}(m-1)}]\\\label{37}
&-&c_{1}(\frac{8\alpha}{(3\alpha-2(m+2))})T^{-1-\frac{8\alpha}{(3\alpha-2(m+2))}}],
\end{eqnarray}
where $\rho$ is given by Eq.(\ref{34}). The anisotropic expansion
measure of anisotropic fluid is
\begin{equation}\label{38}
\frac{\delta-\gamma}{\omega}=\frac{n(m+2)^{2}(2-\frac{2n(m+2)^{2}T^{-\alpha}}
{3\phi_{0}\alpha(m-1)})T^{-2(m+2+\alpha)}}{3\omega(T)}.
\end{equation}
\begin{figure}
\centering \epsfig{file=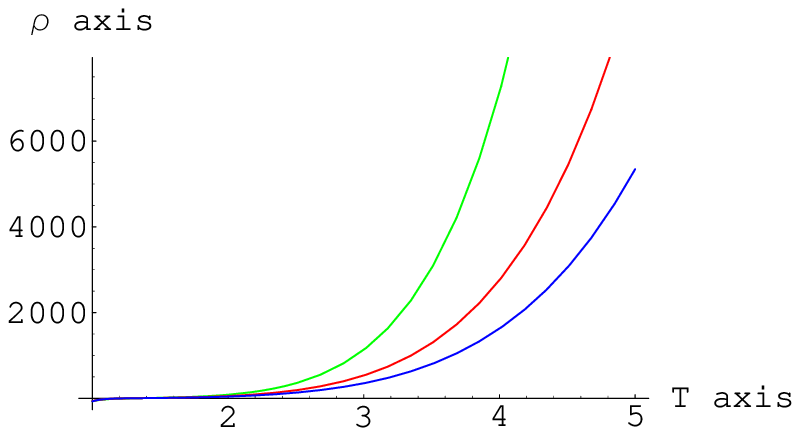,width=.45\linewidth}
\epsfig{file=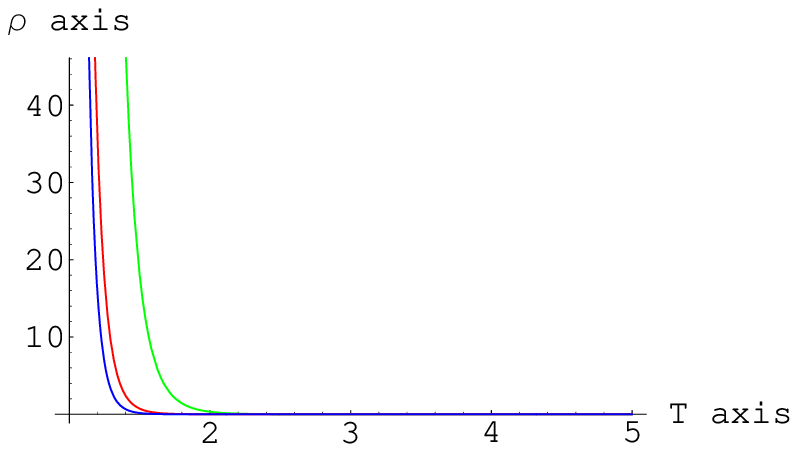,width=.45\linewidth} \caption{Plots
represent energy density $\rho$ versus time T for
$\alpha<\frac{2(m+2)}{3}$ and $\alpha>\frac{2(m+2)}{3}$
respectively. Here $\omega_{0}=1.9,~ n=2$ and $\alpha=1$.}
\end{figure}
\begin{figure}
\centering \epsfig{file=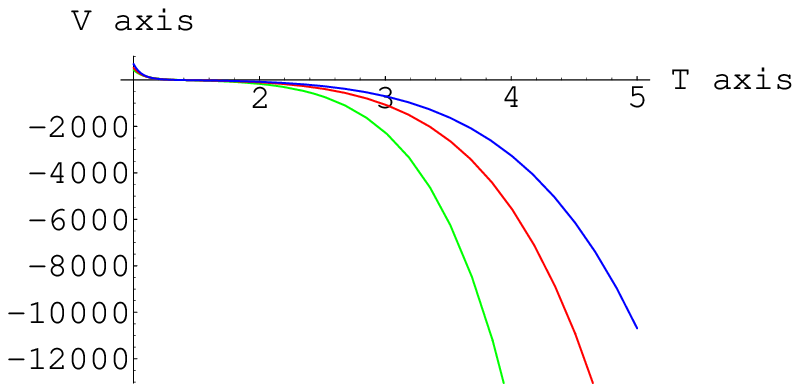,width=.45\linewidth}
\epsfig{file=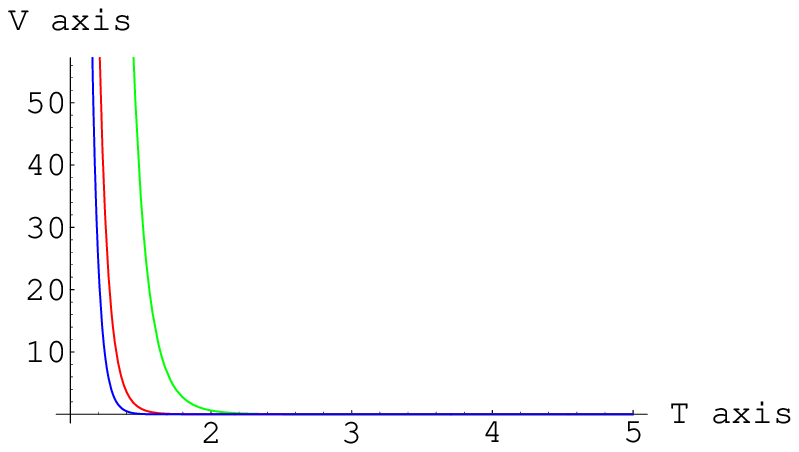,width=.45\linewidth} \caption{The
self-interacting potential $U(T)$ versus time T for
$\alpha<\frac{2(m+2)}{3}$ and $\alpha>\frac{2(m+2)}{3}$. Here
$\omega_{0}=-1.9,~ \beta=2, n=-2,~ \alpha=1$.}
\end{figure}
\begin{figure}
\centering \epsfig{file=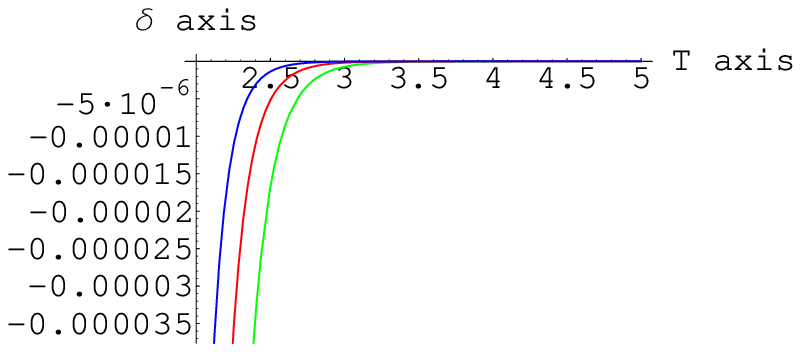,width=.45\linewidth}
\epsfig{file=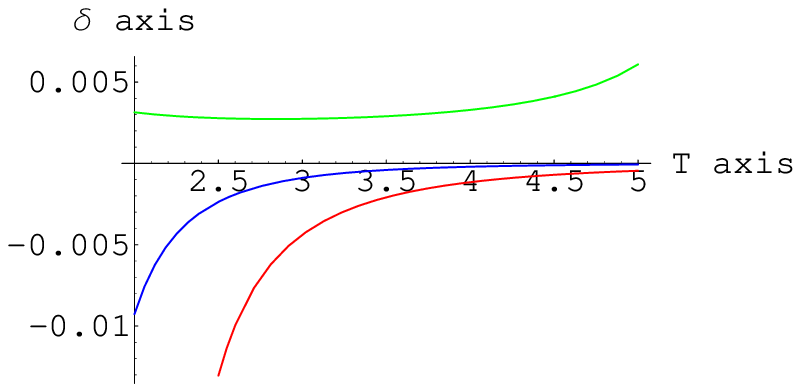,width=.45\linewidth} \caption{The skewness
parameter $\delta(T)$ for $\alpha<\frac{2(m+2)}{3}$ and
$\alpha>\frac{2(m+2)}{3}$ respectively. Here $\omega_{0}=-1.9,~
\alpha=1,~n=2$.}
\end{figure}
\begin{figure}
\centering \epsfig{file=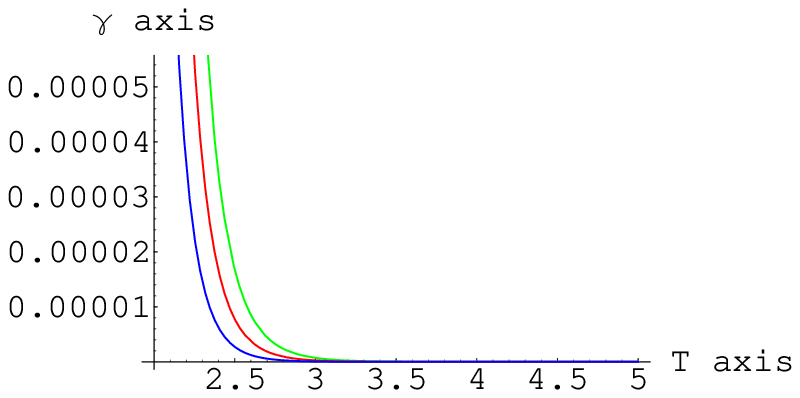,width=.45\linewidth}
\epsfig{file=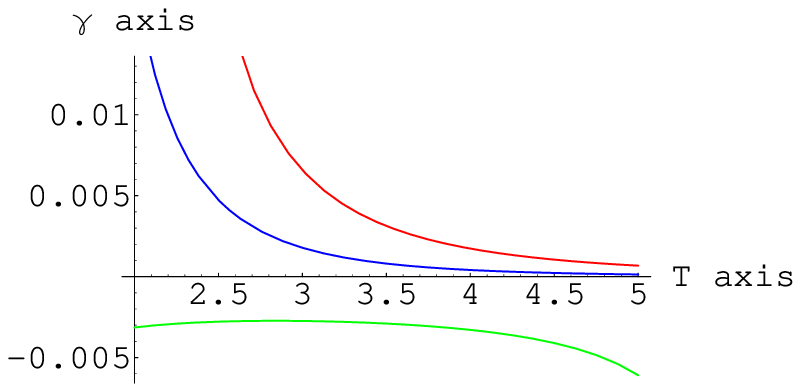,width=.45\linewidth} \caption{The skewness
parameter $\gamma(T)$ for $\alpha<\frac{2(m+2)}{3}$ and
$\alpha>\frac{2(m+2)}{3}$ respectively.}
\end{figure}
\begin{figure}
\centering \epsfig{file=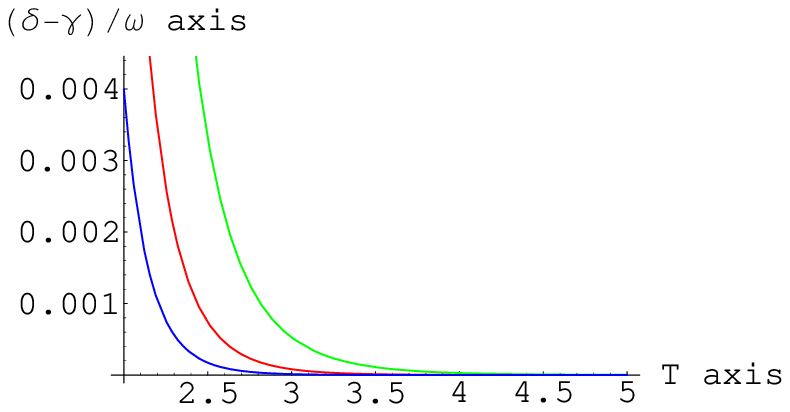,width=.45\linewidth}
\epsfig{file=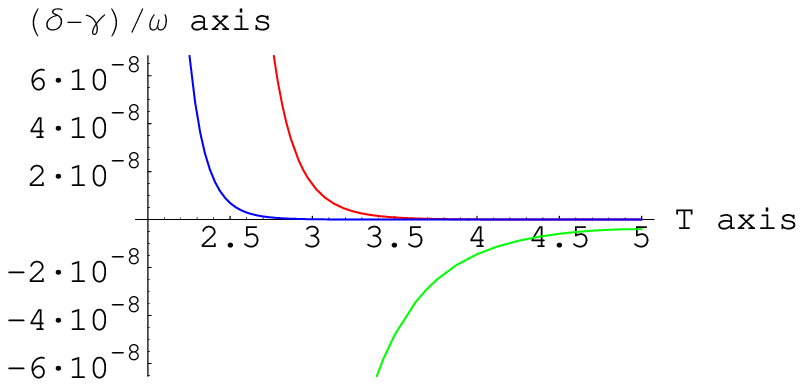,width=.45\linewidth} \caption{The
anisotropic measure of expansion parameter
$(\delta-\gamma)/\omega$ for $\alpha<\frac{2(m+2)}{3}$ and
$\alpha>\frac{2(m+2)}{3}$ respectively.}
\end{figure}

Now we discuss the results for $\alpha<\frac{2(m+2)}{3}$ and
$\alpha>\frac{2(m+2)}{3}$. Figure \textbf{1} indicates that the
energy density is positive. For later times, it decreases and goes
to zero for $\alpha>\frac{2(m+2)}{3}$ while it increases and
approaches to infinity after big bang for $\alpha<\frac{2(m+2)}{3}$.
The self-interacting potential is positive only for
$\alpha>\frac{2(m+2)}{3}$ as shown in Figure \textbf{2} and goes to
zero for later times. Figures \textbf{3} and \textbf{4} show that
the skewness parameters $\delta(T)$ and $\gamma(T)$ turn out to be
finite at $T=0$ and approach to zero in future evolution of the
universe for both cases. The anisotropy measure of expansion for
anisotropic fluid goes to zero for $T\rightarrow\infty$ which shows
that the anisotropic fluid approaches to isotropy for future
evolution of the universe as shown in Figure \textbf{5}. Notice that
all these parameters decrease more rapidly with increasing values of
the parameter $m$.

At the initial epoch with $\alpha<\frac{2(m+2)}{3}$ and
$\alpha>\frac{2(m+2)}{3}$, we obtain
$\omega=-1-\frac{(4+2m+3\alpha)}{(m+2)}$ and
$\omega=-1+\frac{8\alpha}{(m+2)(3\alpha-2(m+2))}$ respectively.
These indicate that for $m>1$ and $\alpha>0$, the universe may be
in phantom region or quintessence region. For later times with
$\alpha<\frac{2(m+2)}{3}$, we have
$\omega=-1+\frac{8\alpha}{(m+2)(3\alpha-2(m+2))}$ and for
$\alpha>\frac{2(m+2)}{3}$, it follows that
\begin{eqnarray}\nonumber
\omega&=&-1-\frac{1}{(m+2)}[[\phi_{0}\alpha(m+2)((3+
2\omega_{0})\alpha(\alpha+m+1)+4(1+2m)\\\nonumber
&+&4\alpha(m+2)-2\omega_{0}\alpha^{2})-(\alpha-2
+\frac{8\alpha(m+2)}{(3\alpha-2(m+2))})\frac{\phi_{0}\alpha^{2}(m+2)}{2}\\\nonumber
&\times&(3+2\omega_{0})]
[-2(\alpha+2m+4)]+\alpha^{2}\phi_{0}\frac{(3+2\omega_{0})(m+2)}{(3\alpha-2(m+2))}
[-(\alpha+2m\\\nonumber
&+&4)][[\phi_{0}\alpha(m+2)((3+2\omega_{0})\alpha(\alpha+m+1)+4(1+2m)
+4\alpha(m+2)\\\nonumber
&-&2\omega_{0}\alpha^{2})-\frac{\phi_{0}\alpha^{2}(m+2)(3+2\omega_{0})}{2}
(\frac{8\alpha(m+2)}{(3\alpha-2(m+2))}+\alpha-2)][2\\\nonumber
&\times&(8\alpha(m+2)-(\alpha+2m
+4)(3\alpha-2(m+2)))^{-1}]\\\nonumber
&+&\alpha^{2}\phi_{0}\frac{(3+2\omega_{0})(m+2)}{(3\alpha-2(m+2))}]^{-1}.
\end{eqnarray}
This also shows that the universe will be in quintessence region or
phantom region for later times depending on the value of the BD
parameter. Thus the model represents accelerated expansion of the
universe.

\section{Magnetized Anisotropic Fluid Model}

In this section, we explore solution of the field equations for
magnetized anisotropic fluid. We take anisotropic fluid with
magnetic field along $z$ axis and assume that there is no electric
field. In this case, the scale factor $A(t)$ is perpendicular to
magnetic field while $B(t)$ is along the field lines. The
magnetized anisotropic fluid is
\begin{equation}\label{40}
T^{\nu}_{\mu}=diag[\rho+\rho_{B}, -p_{x}+\rho_{B},
-p_{y}-\rho_{B}, -p_{z}-\rho_{B}],
\end{equation}
where $\rho_{B}$ represents energy density of the magnetic field.
Using EoS for pressures in $x,~y$ and $z$ directions as in the
anisotropic fluid, Eq.(\ref{40}) can be written as
\begin{equation}\label{41}
T^{\nu}_{\mu}=diag[\rho+\rho_{B}, -(\omega+\delta)+\rho_{B},
-(\omega+\gamma)-\rho_{B}, -(\omega+\gamma)-\rho_{B}],
\end{equation}
where $\delta$ and $\gamma$ are given by Eq.(\ref{26}). For
$\delta=0=\gamma$, Eq.(\ref{41}) corresponds to the
energy-momentum tensor for the magnetized isotropic fluid while it
reduces to the anisotropic fluid for $\rho_{B}=0$. For
$\delta=0=\gamma$ and $\rho_{B}=0$, it represents the isotropic
fluid.

The field equations (\ref{2}) and (\ref{3}) for the model
(\ref{4}) and the energy-momentum tensor (\ref{41}) become
\begin{eqnarray}\label{42}
(2m+1)\frac{\dot{B}^{2}}{B^{2}}&=&\frac{\rho+\rho_{B}}{\phi}+\frac{\omega_{0}}{2}
\frac{\dot{\phi}^{2}}{\phi^{2}}-(m+2)\frac{\dot{B}}{B}\frac{\dot{\phi}}{\phi}
+\frac{U(\phi)}{2\phi},\\\nonumber
2\frac{\ddot{B}}{B}+\frac{\dot{B}^{2}}{B^{2}}&=&-\frac{(\omega+\delta)\rho-\rho_{B}}{\phi}
-\frac{\omega_{0}}{2}
\frac{\dot{\phi}^{2}}{\phi^{2}}-2\frac{\dot{B}}{B}\frac{\dot{\phi}}{\phi}
-\frac{\ddot{\phi}}{\phi}\\\label{43}
&+&\frac{U(\phi)}{2\phi},\\\nonumber
(m+1)\frac{\ddot{B}}{B}+m^{2}\frac{\dot{B}^{2}}{B^{2}}
&=&-\frac{(\omega+\gamma)\rho+\rho_{B}}{\phi}-\frac{\omega_{0}}{2}
\frac{\dot{\phi}^{2}}{\phi^{2}}-\frac{\ddot{\phi}}{\phi}\\\label{44}
&-&(m+1)\frac{\dot{B}}{B}\frac{\dot{\phi}}{\phi}+\frac{U(\phi)}{2\phi},\\\label{45}
\ddot{\phi}+(m+2)\frac{\dot{B}}{B}\dot{\phi}&=&\frac{\rho(1-3\omega)
-\rho(\delta+2\gamma)}{(2\omega_{0}+3)}
-\frac{2U(\phi)-\phi\frac{dU}{d\phi}}{(2\omega_{0}+3)},
\end{eqnarray}
where we have used the condition $A=B^{m}$.

The energy conservation equation for the magnetized anisotropic
fluid yields $\rho_{B}=\frac{\beta}{B^{4}}$ along with
Eq.(\ref{24}). Here $\beta>0$ is an integration constant.
Subtraction of Eq.(\ref{43}) from (\ref{44}) leads to
\begin{equation}\label{46}
2\frac{\ddot{B}}{B}+2(m+1+\alpha)(\frac{\dot{B}^{2}}{B^{2}})
-\frac{2n(m+2)^{2}}{3\phi_{0}(m-1)B^{\alpha}}(\frac{\dot{B}^{2}}{B^{2}})
=-\frac{4\beta}{\phi_{0}(m-1)B^{\alpha+4}}.
\end{equation}
Taking $\dot{B}=f(B)$, this turns out to be
\begin{equation}\label{47}
\frac{df^{*}}{dB}+\frac{2}{B}[(m+1+\alpha)-\frac{nl}{3\phi_{0}(m-1)B^{\alpha}}]f^{*}
=\frac{-4\beta}{\phi_{0}(m-1)B^{\alpha+3}},
\end{equation}
where $f^{*}=f^{2}$ and $l=(m+2)^{2}$ is a positive constant. This
is the first-order linear non-homogeneous differential equation
with variable coefficients whose integrating factor is
$B^{2(m+1+\alpha)}e^{\frac{2nlB^{-\alpha}}{3\phi_{0}\alpha(m-1)}}$.
After some manipulation, the solution becomes
\begin{eqnarray}\nonumber
f^{2}=\dot{B}^{2}&=&\frac{-4\beta
B^{-(\alpha+2)}}{\phi_{0}(\alpha+2m)(m-1)}-\frac{4nl\beta
B^{-2(\alpha+1)}}{3m(m-1)^{2}\phi_{0}^{2}(\alpha+2m)}\\\label{48}
&\times&(1-\frac{2nlB^{-\alpha}}{3\phi_{0}(m-1)\alpha})+c_{2}B^{-2(m+1+\alpha)}
(1-\frac{2nlB^{-\alpha}}{3\phi_{0}(m-1)\alpha}),
\end{eqnarray}
where $c_{2}$ is an integration constant. This can also be written
as
\begin{eqnarray}\nonumber
dt&=&\int[\frac{-4\beta
B^{-(\alpha+2)}}{\phi_{0}(\alpha+2m)(m-1)}-\frac{4nl\beta
B^{-2(\alpha+1)}}{3m(m-1)^{2}\phi_{0}^{2}(\alpha+2m)}\\\nonumber
&\times&(1-\frac{2nlB^{-\alpha}}{3\phi_{0}(m-1)\alpha})+c_{2}B^{-2(m+1+\alpha)}
(1-\frac{2nlB^{-\alpha}}{3\phi_{0}(m-1)\alpha})]^{-1/2}dB.
\end{eqnarray}
By taking $B=T,~x=X,~y=Y$ and $z=Z$ and using Eq.(\ref{48}), BI
spacetime turns out to be
\begin{eqnarray}\nonumber
ds^{2}&=&[\frac{-4\beta
T^{-(\alpha+2)}}{\phi_{0}(\alpha+2m)(m-1)}-\frac{4nl\beta
T^{-2(\alpha+1)}}{3m(m-1)^{2}\phi_{0}^{2}(\alpha+2m)}\\\nonumber
&\times&(1-\frac{2nlT^{-\alpha}}{3\phi_{0}(m-1)\alpha})+c_{2}T^{-2(m+1+\alpha)}
(1-\frac{2nlT^{-\alpha}}{3\phi_{0}(m-1)\alpha})]^{-1}dT^{2}\\\label{49}
&-&T^{2m}dX^{2}-T^{2}(dY^{2}+dZ^{2}).
\end{eqnarray}

Now we discuss some physical features of this model. Since at
$T=0$, the scale factors will be zero, the model shows point type
singularity \cite{18, 22}. The corresponding mean and directional
Hubble parameters are
\begin{eqnarray}\nonumber
H_{x}=mH_{y}&=&m[\frac{-4\beta
T^{-(\alpha+4)}}{\phi_{0}(\alpha+2m)(m-1)}-\frac{4nl\beta
T^{-2(\alpha+2)}}{3m(m-1)^{2}\phi_{0}^{2}(\alpha+2m)}\\\nonumber
&\times&(1-\frac{2nlT^{-\alpha}}{3\phi_{0}(m-1)\alpha})+c_{2}T^{-2(m+2+\alpha)}
(1-\frac{2nlT^{-\alpha}}{3\phi_{0}(m-1)\alpha})]^{1/2},\\\label{50}
\\\nonumber
H&=&\frac{(m+2)}{3}[\frac{-4\beta
T^{-(\alpha+4)}}{\phi_{0}(\alpha+2m)(m-1)}-\frac{4nl\beta
T^{-2(\alpha+2)}}{3m(m-1)^{2}\phi_{0}^{2}(\alpha+2m)}\\\nonumber
&\times&(1-\frac{2nlT^{-\alpha}}{3\phi_{0}(m-1)\alpha})+c_{2}T^{-2(m+2+\alpha)}
(1-\frac{2nlT^{-\alpha}}{3\phi_{0}(m-1)\alpha})]^{1/2}.\\\label{51}
\end{eqnarray}
Since $\alpha,~m>0~(m\neq1)$, these parameters increase with the
decrease in $T$ and approach to zero as $T\rightarrow\infty$. Also,
these parameters take infinitely large values at $T=0$. The
remaining parameters are given by
\begin{eqnarray}\nonumber
\Theta&=&3H=(m+2)[\frac{-4\beta
T^{-(\alpha+4)}}{\phi_{0}(\alpha+2m)(m-1)}-\frac{4nl\beta
T^{-2(\alpha+2)}}{3m(m-1)^{2}\phi_{0}^{2}(\alpha+2m)}\\\label{53}
&\times&(1-\frac{2nlT^{-\alpha}}{3\phi_{0}(m-1)\alpha})+c_{2}T^{-2(m+2+\alpha)}
(1-\frac{2nlT^{-\alpha}}{3\phi_{0}(m-1)\alpha})]^{1/2},\\\nonumber
\sigma^2&=&\frac{(m-1)^2}{3}[\frac{-4\beta
T^{-(\alpha+4)}}{\phi_{0}(\alpha+2m)(m-1)}-\frac{4nl\beta
T^{-2(\alpha+2)}}{3m(m-1)^{2}\phi_{0}^{2}(\alpha+2m)}\\\label{54}
&\times&(1-\frac{2nlT^{-\alpha}}{3\phi_{0}(m-1)\alpha})+c_{2}T^{-2(m+2+\alpha)}
(1-\frac{2nlT^{-\alpha}}{3\phi_{0}(m-1)\alpha})],\\\nonumber
q&=&-(1-\frac{3}{(m+2)})-\frac{3}{2(m+2)}[\frac{-4\beta
T^{-(\alpha+4)}}{\phi_{0}(\alpha+2m)(m-1)}-\frac{4nl\beta
}{3m(m-1)^{2}}\\\nonumber
&\times&\frac{T^{-2(\alpha+2)}}{\phi_{0}^{2}(\alpha+2m)}(1-\frac{2nlT^{-\alpha}}
{3\phi_{0}(m-1)\alpha})+c_{2}T^{-2(m+2+\alpha)}
(1-\frac{2nl\phi_{0}^{-1}}{3(m-1)\alpha}\\\nonumber&\times&T^{-\alpha})]^{-1/2}(\frac{-4\beta
B^{-(\alpha+2)}}{\phi_{0}(\alpha+2m)(m-1)}-\frac{4nl\beta
B^{-2(\alpha+1)}}{3m(m-1)^{2}\phi_{0}^{2}(\alpha+2m)}\\\nonumber
&\times&(1-\frac{2nlB^{-\alpha}}{3\phi_{0}(m-1)\alpha})+c_{2}B^{-2(m+1+\alpha)}
(1-\frac{2nlB^{-\alpha}}{3\phi_{0}(m-1)\alpha}))^{-1/2}
\end{eqnarray}
\begin{eqnarray}\nonumber
&\times&(\frac{4(\alpha+2)\beta
B^{-(\alpha+3)}}{\phi_{0}(\alpha+2m)(m-1)}+\frac{8(\alpha+1)nl\beta
B^{-(2\alpha+3)}}{3m(m-1)^{2}\phi_{0}^{2}(\alpha+2m)}\\\nonumber
&\times&(1-\frac{2nlB^{-\alpha}}{3\phi_{0}(m-1)\alpha})-\frac{4nl\beta
B^{-2(\alpha+1)}}{3m(m-1)^{2}\phi_{0}^{2}(\alpha+2m)}(\frac{2nl\alpha
B^{-(\alpha+1)}}{3\phi_{0}(m-1)\alpha})\\\nonumber
&-&2c_{2}(m+1+\alpha)B^{-(2m+3+2\alpha)}
(1-\frac{2nlB^{-\alpha}}{3\phi_{0}(m-1)\alpha})\\\nonumber
&+&c_{2}B^{-2(m+1+\alpha)}
(\frac{2nl\alpha B^{-(\alpha+1)}}{3\phi_{0}(m-1)\alpha})).
\end{eqnarray}
In this case, the volume of the universe and anisotropic parameter
of expansion turn out to be the same as in anisotropic case. For
initial time, the expansion and shear scalar become infinite while
for later times, these decrease to zero. The deceleration parameter
turns out to be a dynamical quantity. It can be negative for
appropriate values of the constant parameters e.g., it becomes a
negative for later times with $m>1$. Notice that the expansion
scalar, shear scalar and Hubble parameters are decreased by the
component of magnetic field.

We solve Eqs.(\ref{42}) and (\ref{45}) simultaneously to obtain
density $\rho$ and the self-interacting potential $U(\phi)\approx
U(T)$. The density is
\begin{eqnarray}\nonumber
\rho(T)&=&\phi_{0}(1+2m+\alpha(m+2)
-\frac{\omega_{0}\alpha^{2}}{2})[\frac{-4\beta
T^{-4}}{(\alpha+2m)(m-1)\phi_{0}}\\\nonumber&-&\frac{4nl\beta
}{3m(m-1)^{2}}\frac{T^{-(\alpha+4)}}{(\alpha+2m)\phi_{0}^{2}}(1
-\frac{2nlT^{-\alpha}}{3\phi_{0}(m-1)\alpha})\\\label{55}&+&c_{2}T^{-(4+2m+\alpha)}(1
-\frac{2nlT^{-\alpha}}{3\phi_{0}(m-1)\alpha})]-\frac{\beta}{T^{4}}-\frac{U(T)}{2},
\end{eqnarray}
where $U(T)$ is
\begin{eqnarray}\nonumber
U(T)&=&[((2\omega_{0}+3)(\alpha+m+1)\alpha^{2}-4\alpha(1+2m+\alpha(m+2)
-\frac{\omega_{0}\alpha^{2}}{2}))2(m\\\nonumber
&+&2)+\alpha^{2}(m+2)(3+2\omega_{0})(\frac{8\alpha(m+2)}
{2(m+2)-3\alpha}-(\alpha-2))-6\alpha(1+2m \\\nonumber
&+&\alpha(m+2)
-\frac{\omega_{0}\alpha^{2}}{2})\frac{8\alpha(m+2)}{2(m+2)-3\alpha}][4\beta
T^{-4}[(\alpha+2m)(m-1)\phi_{0}(8\alpha\\\nonumber
&\times&(m+2)+4(2(m+2)-3\alpha))]^{-1}-\frac{4\beta
nl}{3m(\alpha+2m)(m-1)^{2}\phi_{0}^{2}}
\end{eqnarray}
\begin{eqnarray}\nonumber
&\times&(-T^{-(\alpha+4)}[(8\alpha(m+2)+(\alpha+
4)(2(m+2)-3\alpha))]^{-1}+\frac{2}{3}nl\phi_{0}^{-1}(m\\\nonumber
&-&1)^{-1} T^{-(2\alpha+4)}(8\alpha^{2}(m+2)+2\alpha(\alpha+
2)(2(m+2)-3\alpha))^{-1}-c_{2}(8\alpha\\\nonumber
&\times&(m+2)-(\alpha+
2m+4)(2(m+2)-3\alpha))^{-1}T^{-(4+2m+\alpha)}+\frac{2}{3
}nl\phi_{0}^{-1}\\\nonumber
&\times&(m-1)^{-1}\alpha
T^{-(2\alpha+2m+4)}(8\alpha(m+2)+2(\alpha+m+2)(2(m+2)\\\nonumber
&-&3\alpha))^{-1}+\frac{\alpha^{2}(2\omega_{0}+3)}{(2(m+2)-3\alpha)}[\frac{-4(m+2)\beta
T^{-4}}{(\alpha+2m)(m-1)\phi_{0}}-\frac{4(m+2)}{3m(m-1)^{2}}
\\\nonumber
&\times&\frac{\beta
nl}{(\alpha+2m)}(T^{-(\alpha+4)}-\frac{2nlT^{-2(\alpha
+2)\phi_{0}^{-1}}}{3\alpha(m-1)})
+c_{2}(m+2)(T^{-(\alpha+2m+4)}\\\nonumber
&-&\frac{2nl\phi_{0}^{-1}}{3\alpha(m-1)}T^{-(2\alpha+2m+4)}]
+\frac{4n(m+2)^{2}(1-m)\alpha}{3} [\frac{4\beta
T^{-(\alpha+4)}\phi_{0}^{-1}}{(\alpha+2m)(m-1)}\\\nonumber
&\times&(8\alpha(m+2)+(\alpha+4) (2(m+2)-3\alpha))^{-1}
-\frac{4\beta nl\phi_{0}^{-2}}{3m(\alpha+2m)(m-1)^{2}}\\\nonumber
&\times&(-T^{-(2\alpha+4)}(8\alpha(m+2)+(2\alpha
+4)(2(m+2)-3\alpha))^{-1}\\\nonumber
&+&\frac{2nlT^{-(3\alpha+4)}}{3\phi_{0}(m-1)\alpha}(8\alpha(m+2)
+(3\alpha+4)(2(m+2)-3\alpha))^{-1}\\\nonumber
&+&c_{2}(-T^{-(4+2m+2\alpha)}(8\alpha(m+2)+(2\alpha
+2m+4)(2(m+2)-3\alpha))^{-1}\\\nonumber
&+&\frac{2nlT^{-(3\alpha+2m+4)}}
{3\phi_{0}(m-1)\alpha}(8\alpha(m+2)+(3\alpha+2m+
4)(2(m+2)-3\alpha))^{-1})\\\nonumber
&-&8\alpha\beta(m+2)T^{-4}(8\alpha(m+2)+4(2(m+2)-3\alpha))^{-1}
+6\alpha(2(m+2)\\\nonumber
&-&3\alpha)_{-1}[\beta
T^{-4}-8\beta\alpha(m+2)T^{-4}(8\alpha(m+2)+4(2(m+2)-3\alpha))^{-1}]\\\nonumber
&-&\frac{6\alpha}{(2(m+2)-3\alpha)}(1+2m+\alpha(m+2)
-\frac{\omega_{0}\alpha^{2}}{2})[\frac{-4\beta
T^{-(\alpha+3)}}{(\alpha+2m)(m-1)}\\\nonumber
&-&\frac{4\beta
nl}{3m(m-1)^{2}(\alpha+2m)}(T^{-(2\alpha+3)}-\frac{2nlT^{-3(\alpha+1)}}{3\phi_{0}(m-1)\alpha})
+c_{2}(T^{-(2\alpha+2m+3)}\\\label{56}
&-&\frac{2nlT^{-(3\alpha+2m+3)}}{3\alpha(m-1)\phi_{0}})]
+c_{3}T^{\frac{8\alpha(m+2)}{2(m+2)-3\alpha}},
\end{eqnarray}
where $c_{3}$ is an integration constant. Figure \textbf{6}
indicates that the energy density is positive and decreases after
big bang but it increases and approaches to infinity for later
times with $\alpha<\frac{2(m+2)}{3}$. The energy density is
positive but decreases to zero for any positive value of the
parameter satisfying $\alpha>\frac{2(m+2)}{3}$ as shown in Figure
\textbf{6}. At the initial epoch, there is infinite energy density
in both cases as shown in Figure \textbf{6}. Figures \textbf{7}
indicate that the self-interacting potential remains positive in
both cases ($\alpha<\frac{2(m+2)}{3}$ and
$\alpha>\frac{2(m+2)}{3}$). The corresponding skewness parameters
turn out to be
\begin{eqnarray}\nonumber
\delta(T)&=&\frac{2n(m+2)}{3\rho}[\frac{-4\beta
T^{-(\alpha+4)}}{(\alpha+2m)(m-1)\phi_{0}}-\frac{4nl\beta
T^{-2(\alpha+2)}}{3m(m-1)^{2}(\alpha+2m)\phi_{0}^{2}}\\\label{57}
&\times&(1-\frac{2nlT^{-\alpha}}{3\phi_{0}(m-1)\alpha})+c_{2}T^{-2(2+m+\alpha)}(1
-\frac{2nlT^{-\alpha}}{3\phi_{0}(m-1)\alpha})], \\\nonumber
\gamma(T)&=&\frac{-nm(m+2)}{3\rho}[\frac{-4\beta
T^{-(\alpha+4)}}{(\alpha+2m)(m-1)\phi_{0}}-\frac{4nl\beta
T^{-2(\alpha+2)}}{3m(m-1)^{2}(\alpha+2m)\phi_{0}^{2}}
\\\label{58}&\times&(1-\frac{2nlT^{-\alpha}}{3\phi_{0}(m-1)\alpha})+c_{2}T^{-2(2+m+\alpha)}(1
-\frac{2nlT^{-\alpha}}{3\phi_{0}(m-1)\alpha})],
\end{eqnarray}
where $\rho$ is given by Eq.(\ref{55}). Figures \textbf{8} and
\textbf{9} indicate that the deviation parameters become finite at
$T=0$. For later times, these parameters converge to zero in both
cases. From Eqs.(\ref{24}) and (\ref{25}), the deviation free EoS
parameter $\omega$ can be written as
\begin{equation}\label{59}
\omega(T)=-1-\frac{B}{\rho(m+2)}\frac{d\rho}{dB}.
\end{equation}

The anisotropy measure of anisotropic fluid,
$\frac{\delta-\gamma}{\omega}$, for the model (\ref{49}) takes the
form
\begin{eqnarray}\nonumber
\frac{\delta-\gamma}{\omega}&=&\frac{n(m+2)^{2}}{3\omega(T)}[\frac{-4\beta
T^{-(\alpha+4)}}{(\alpha+2m)(m-1)\phi_{0}}-\frac{4nl\beta
T^{-2(\alpha+2)}}{3m(m-1)^{2}(\alpha+2m)\phi_{0}^{2}}\\\label{60}
&\times&(1-\frac{2nlT^{-\alpha}}{3\phi_{0}(m-1)\alpha})+c_{2}T^{-2(2+m+\alpha)}(1
-\frac{2nlT^{-\alpha}}{3\phi_{0}(m-1)\alpha})].
\end{eqnarray}
Its behavior is shown in Figure \textbf{10}.
\begin{figure}
\centering \epsfig{file=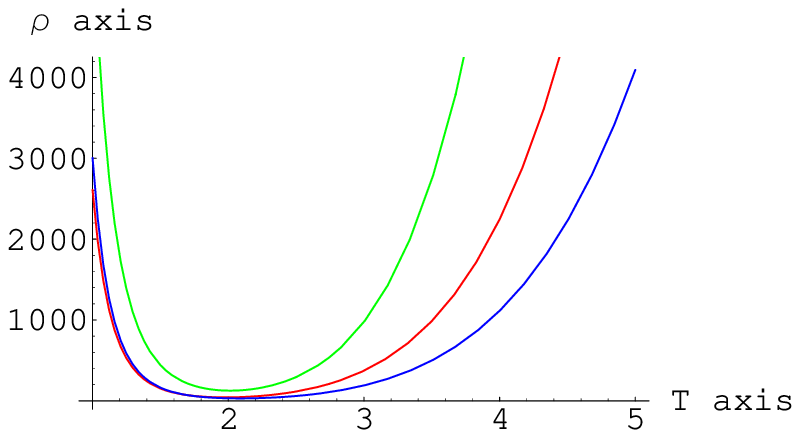,width=.45\linewidth}
\epsfig{file=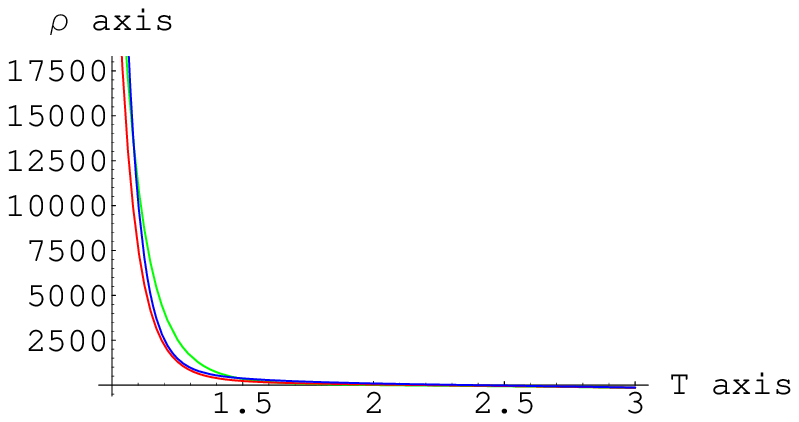,width=.45\linewidth} \caption{Plots
represent the energy density $\rho(T)$ versus time T for
$\alpha<\frac{2(m+2)}{3}$ and $\alpha>\frac{2(m+2)}{3}$
respectively. Here $\omega_{0}=-1.9,~ \beta=2,~n=-2$. Green, red
and blue lines show the graphs for $m=2,3,4$ respectively.}
\end{figure}
\begin{figure}
\centering \epsfig{file=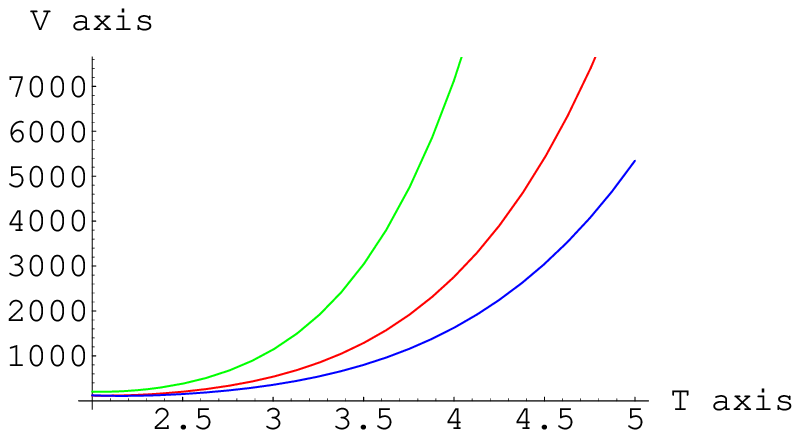,width=.45\linewidth}
\epsfig{file=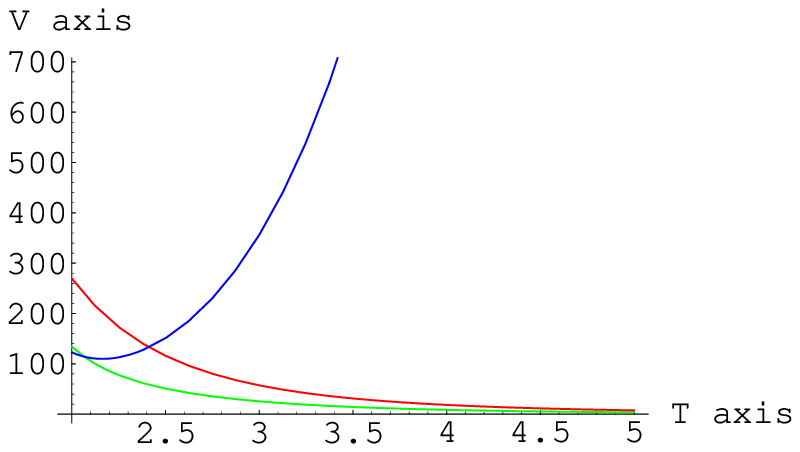,width=.45\linewidth} \caption{Plots show the
self-interacting potential $U(T)$ versus time T for
$\alpha<\frac{2(m+2)}{3}$ and $\alpha>\frac{2(m+2)}{3}$
respectively.}
\end{figure}
\begin{figure}
\centering \epsfig{file=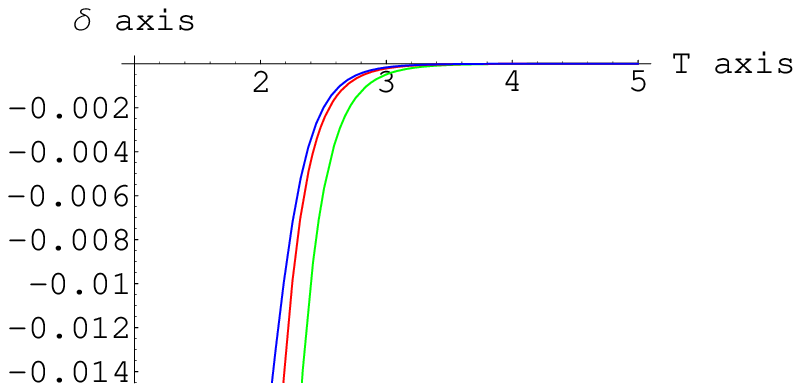,width=.45\linewidth}
\epsfig{file=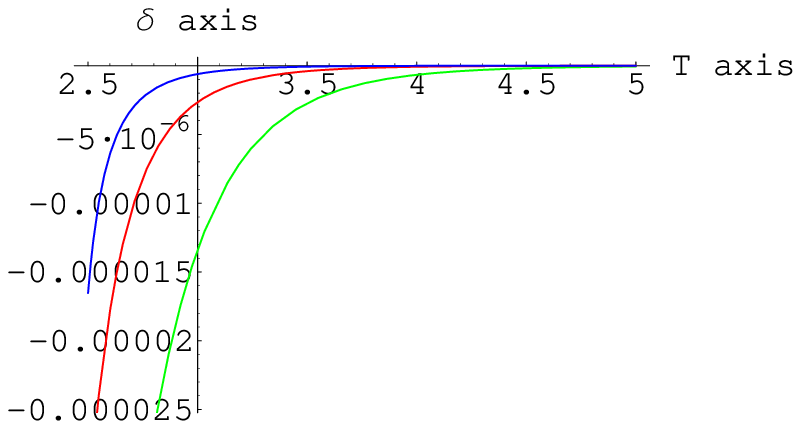,width=.45\linewidth} \caption{The deviation
parameter $\delta(T)$ versus time T for $\alpha<\frac{2(m+2)}{3}$
and $\alpha>\frac{2(m+2)}{3}$ respectively. Here
$\omega_{0}=-1.9,~ \beta=2$ and $n=-2$.}
\end{figure}
For initial epoch, this is finite while it goes to zero for the
future evolution of the universe in both cases. This indicates
that the anisotropic fluid approaches to isotropy for later times.
When $T\longrightarrow 0$ and $\alpha<\frac{2(m+2)}{3}$, we obtain
$\omega=-1-\frac{(4+2m+3\alpha)}{(m+2)}$ which shows that the
universe model will be in phantom region at initial epoch. For
$\alpha>\frac{2(m+2)}{3}$, we obtain
$\omega=-1+\frac{8\alpha}{(m+2)(3\alpha-2(m+2))}$ which shows that
the universe model will be in quintessence region at initial
epoch. For later times with $\alpha<\frac{2(m+2)}{3}$, it follows
that $\omega=-1+\frac{8\alpha}{(m+2)(3\alpha-2(m+2))}$ showing
that the universe may be in quintessence region. When
$\alpha>\frac{2(m+2)}{3}$, the EoS parameter depends on the
component of magnetic field and BD parameter indicating that the
universe will be in phantom or quintessence region for appropriate
values of the constant parameters. Thus, in each case, the model
shows the accelerated expansion of the universe.
\begin{figure}
\centering \epsfig{file=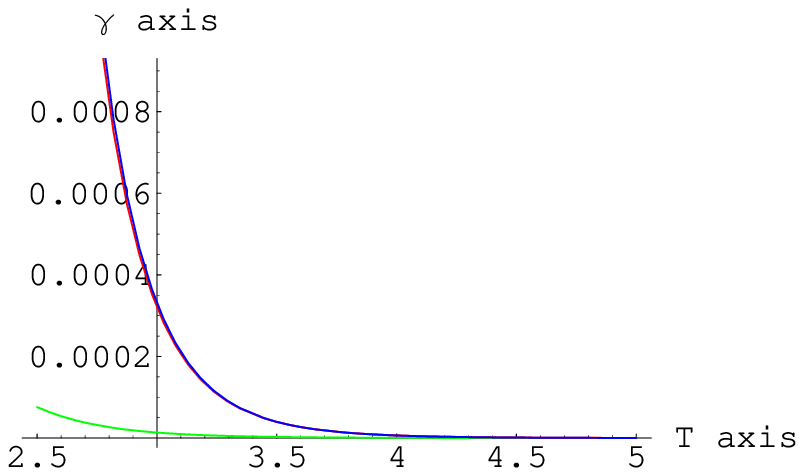,width=.45\linewidth}
\epsfig{file=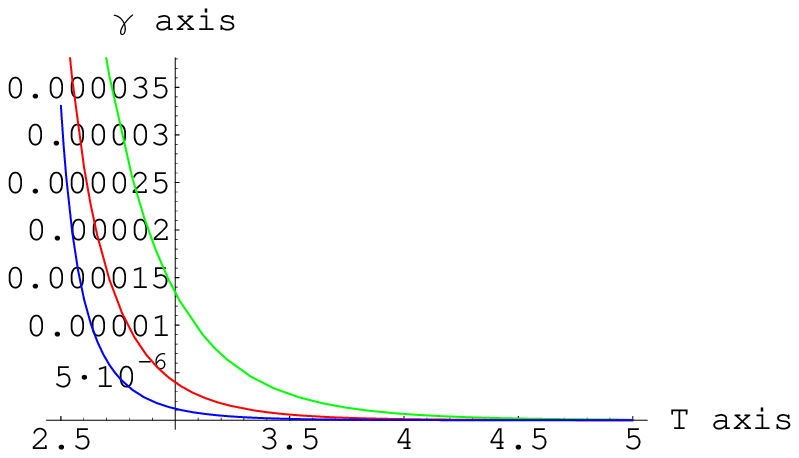,width=.45\linewidth} \caption{The deviation
parameter $\gamma(T)$ versus time T for $\alpha<\frac{2(m+2)}{3}$
and $\alpha>\frac{2(m+2)}{3}$ respectively.}
\end{figure}
\begin{figure} \centering
\epsfig{file=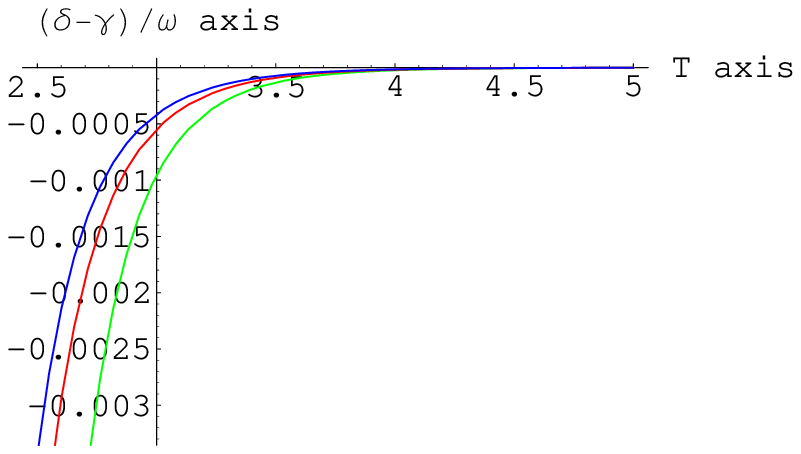,width=.45\linewidth}
\epsfig{file=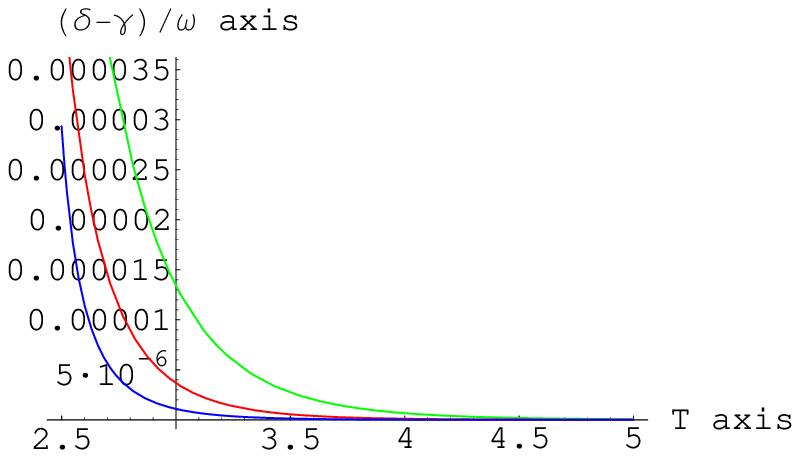,width=.45\linewidth} \caption{Anisotropic
measure of expansion $\frac{\delta-\gamma}{\omega}$ versus time T
is shown for $\alpha<\frac{2(m+2)}{3}$ and
$\alpha>\frac{2(m+2)}{3}$ respectively.}
\end{figure}

Now we investigate a special case when $m=1$. The scale factors
become $A(t)=B(t)=a(t)$ and the model turns out to be the FRW
universe model
\begin{equation*}
ds^{2}=dt^{2}-a(t)^{2}(dx^{2}+dy^{2}+dz^{2}).
\end{equation*}
Equation (\ref{46}) yields
\begin{equation}\label{61}
a(t)=\sqrt{2}(\sqrt{\frac{2\beta}{3n}}t+c_{4})^{1/2},
\end{equation}
where $c_{4}$ is an integration constant. The expansion scalar
$\Theta$ turns out to be
\begin{equation}\nonumber
\Theta=3H=3(\frac{\dot{a}}{a})
=3\sqrt{\frac{2\beta}{3n}}(\sqrt{\frac{2\beta}{3n}}t+c_{4})^{-1}.
\end{equation}
This shows that the Hubble parameter and the expansion scalar are
constant at earlier time. As time increases, both of these
parameters decrease indicating expanding universe in its earlier
time. From Eqs.(\ref{42}) and (\ref{45}), the energy density
$\rho(t)$  and the self-interacting potential $U(t)$ are
\begin{eqnarray}\nonumber
\rho(t)&=&S_{1}[\sqrt{\frac{2\beta}{3n}}t+c_{4}]^{\frac{\alpha-2}{2}}
+S_{2}[\sqrt{\frac{2\beta}{3n}}t+c_{4}]^{\frac{\alpha-4}{2}}
+c_{5}[2(\sqrt{\frac{2\beta}{3n}}t\\\nonumber
&+&c_{4})]^{\frac{4\alpha}{(2-\alpha)}}
-\frac{\beta}{4}[\sqrt{\frac{2\beta}{3n}}t+c_{4}]^{-2},\\\nonumber
U(t)&=&-[2\phi_{0}\alpha((3+2\omega_{0})\alpha(\alpha+2)
-12(1+\alpha)+2\omega_{0}\alpha^{2})+\frac{\alpha^{2}
(3+2\omega_{0})}{(2-\alpha)}\\\nonumber
&\times&\phi_{0}(\alpha+2)^{2}-\frac{16\phi_{0}\alpha^{2}}{(2-\alpha)}(3(1+\alpha)
-\frac{\omega_{0}\alpha^{2}}{2})]
[\sqrt{\frac{2\beta}{3n}}t+c_{4}]^{\frac{\alpha-2}{2}}(\frac{2\beta}{3n})^{3/2}\\\nonumber
&\times&(\alpha+2)^{-2}+[\frac{2^{(\alpha-2)/2}\phi_{0}\alpha^{2}(3+2\omega_{0})\beta}{3n(2-\alpha)}
-\frac{4\alpha\phi_{0}\beta
2^{\alpha/2}}{3n(2-\alpha)}(3(1+\alpha)\\\nonumber
&-&\frac{\omega_{0}\alpha^{2}}{2})]
[\sqrt{\frac{2\beta}{3n}}t+c_{4}]^{\frac{\alpha-4}{2}}+c_{5}[2(\sqrt{\frac{2\beta}{3n}}t
+c_{4})]^{\frac{4\alpha}{(2-\alpha)}}.
\end{eqnarray}
Here $c_{5}$ is an integration constant and the constants $S_{1}$
and $S_{2}$ are given by
\begin{eqnarray}\nonumber
S_{1}&=&-(1/2)[2\phi_{0}\alpha((3+2\omega_{0})\alpha(\alpha+2)-
12(1+\alpha)+2\omega_{0}\alpha^{2})\\\nonumber &+&\frac{\alpha^{2}
(3+2\omega_{0})\phi_{0}}{(2-\alpha)}(\alpha+2)^{2}
-\frac{16\phi_{0}\alpha^{2}}{(2-\alpha)}(3(1+\alpha)-\frac{\omega_{0}\alpha^{2}}{2})]\\\nonumber
&\times&(\frac{2\beta}{3n})^{3/2}(\alpha+2)^{-2},\\\nonumber
S_{2}&=&[\frac{2^{(\alpha-2)/2}\phi_{0}\alpha^{2}(3+2\omega_{0})\beta}{6n(2-\alpha)}
-\frac{4\alpha\phi_{0}\beta2^{\alpha/2}}{6n(2-\alpha)}
(3(1+\alpha)-\frac{\omega_{0}\alpha^{2}}{2})\\\nonumber
&+&\phi_{0}(3(1+\alpha)-\frac{\omega_{0}\alpha^{2}}{2})\frac{2^{(\alpha-2)/2}\beta}{3n}].
\end{eqnarray}

Some other parameters are
\begin{eqnarray}\nonumber
\delta(t)&=&\frac{4\beta}{3\rho}(\sqrt{\frac{2\beta}{3n}}t+c_{4})^{-2},\quad
\gamma(t)=\frac{-2\beta}{3\rho}(\sqrt{\frac{2\beta}{3n}}t+c_{4})^{-2},
\end{eqnarray}
\begin{eqnarray}\nonumber
\omega(t)&=&-1-[\frac{S_{1}(\alpha-2)}{2}(\sqrt{\frac{2\beta}{3n}}t+c_{4})^{(\alpha-4)/2}
+\frac{S_{2}(\alpha-4)}{2}(\sqrt{\frac{2\beta}{3n}}t\\\nonumber
&+&c_{4})^{(\alpha-6)/2}+c_{5}2^{4\alpha/(2-\alpha)}(\frac{4\alpha}{(2-\alpha)}
\sqrt{\frac{2\beta}{3n}}t+c_{4})^{(5\alpha-2)/(2-\alpha)}\\\nonumber
&+&\frac{\beta}{2}(\sqrt{\frac{2\beta}{3n}}t+c_{4})^{-3}]
[3S_{1}[\sqrt{\frac{2\beta}{3n}}t+c_{4}]^{\frac{\alpha-4}{2}}+3S_{2}[\sqrt{\frac{2\beta}{3n}}
t+c_{4}]^{\frac{\alpha-6}{2}}\\\nonumber
&+&3c_{5}2^{4\alpha/(2-\alpha)}[2(\sqrt{\frac{2\beta}{3n}}t
+c_{4})]^{\frac{(5\alpha-2)}{(2-\alpha)}}
-\frac{\beta}{4}[\sqrt{\frac{2\beta}{3n}}t+c_{4}]^{-3}]^{-1},\\\nonumber
\frac{\delta-\gamma}{\omega}&=&\frac{2\beta(\sqrt{\frac{2\beta}{3n}}t+c_{4})^{-2}}{\omega(t)},
\end{eqnarray}
where $\alpha\neq2$. We discuss two cases: $0<\alpha<2$ and
$\alpha>2$. Clearly, the energy density is constant at initial
epoch and approaches to infinity for later times in both cases.
The anisotropy parameters are constant at initial epoch and go to
zero for later times. Likewise the anisotropy measure of expansion
of anisotropic fluid $\frac{\delta-\gamma}{\omega}$ approaches to
isotropy for later times. The anisotropy parameter of expansion is
zero as $m=1$. At the initial epoch, the deviation free EoS
parameter shows that the universe may be in quintessence region by
choosing appropriate values of the constants in both cases. For
later times with $0<\alpha<2$, we obtain
$\omega(t)=-1-\frac{4\alpha}{3(2-\alpha)}$ which indicates that
the universe will be in phantom region. For $\alpha>2$, it follows
that $\omega(t)=-1-\frac{(\alpha-2)}{6}$, which also shows that
the universe will be in phantom region for future evolution.

\section{Summary and Discussion}

In this paper, we have constructed the BI universe models in BD
theory of gravity with perfect, anisotropic and magnetized
anisotropic fluids. We have constructed exact solutions in each
case. For anisotropic and anisotropic magnetized fluid models, the
physical behavior of the energy density, self-interacting
potential, skewness parameters and anisotropy parameter of
expansion of anisotropy fluid have been plotted for non-zero value
of $n$ with $\alpha<\frac{2(m+2)}{3}$ and
$\alpha>\frac{2(m+2)}{3}$. The results are summarized as follows.
\begin{itemize}
\item In the case of anisotropic as well as magnetized anisotropic fluids,
the skewness parameters and anisotropic measure of expansion of
anisotropic fluid go to zero indicating the isotropic behavior of
the fluid for the future evolution of the universe. This result
coincides with those already available in literature for Bianchi
type III model in $f(R)$ theory \cite{11} and Bianchi type
$(VI)_{0}$ model in GR \cite{23}.
\item In each case, the energy density remains positive. All the figures
indicate that energy density increases after big bang and approaches
to infinity for later times with $\alpha<\frac{2(m+2)}{3}$ in both
anisotropic as well as magnetized anisotropic fluids. For
$\alpha>\frac{2(m+2)}{3}$, it decreases and goes to zero in both
cases.
\item For anisotropic fluid, the self interacting potential
is positive only for $\alpha>\frac{2(m+2)}{3}$ and decreases to zero
for later times while for magnetized anisotropic fluid, it remains
positive in both cases.
\item All the physical parameters $H,~H_{x},H_{y},~\Theta$ and
$\sigma$ increase with the decrease in $T$ and go to zero as
$T\rightarrow\infty$. These parameters take infinitely large values
at $T=0$. In contrast to the perfect fluid, the deceleration
parameter for anisotropic fluids is a dynamical quantity and can be
negative for the appropriate choice of constant parameters, in
particular for later times with $m>1$. This corresponds to
accelerated expansion of the universe.
\item In the
anisotropic magnetized fluid, all the physical parameters are
reduced by the component of magnetic field with $n>0$.
\item The deviation free EoS parameters indicate that the universe may
be in quintessence or phantom region at initial epoch as well as
for later times for an appropriate values of the constant
parameters in all cases. Thus the models represent the accelerated
expansion of the universe.
\item The anisotropy parameter of expansion is constant (vanishes for
$m=1$) indicating the model does not isotropize for later times in
all cases.
\item A special case $m=1$ for the magnetized
anisotropic fluid has also been discussed which yields FRW
universe model. In this case, the deviation free EoS parameter
indicates that at initial epoch, the universe may be in
quintessence region while for the future evolution, it will be in
phantom region.
\end{itemize}
It would be interesting to construct exact solutions in the
presence of anisotropic fluid for other Bianchi models in BD
theory.

\end{document}